\documentclass{llncs}
\usepackage{graphicx}

\begin{document}

\title{Hidden Quantum Markov Models and \\ Open Quantum Systems with \\ Instantaneous Feedback}
\author{Lewis A. Clark\inst{1} \and Wei Huang\inst{2} \and Thomas M. Barlow\inst{1} \and Almut Beige\inst{1}}
\institute{The School of Physics and Astronomy, University of Leeds, Leeds, LS2 9JT, United Kingdom
\and 20 Dover Drive Singapore, Singapore University of Technology \& Design, Singapore, 138682, Singapore}

\maketitle

\begin{abstract}
Hidden Markov Models are widely used in classical computer science to model stochastic processes with a wide range of applications. This paper concerns the quantum analogues of these machines --- so-called Hidden Quantum Markov Models (HQMMs). Using the properties of Quantum Physics, HQMMs are able to generate more complex random output sequences than their classical counterparts, even when using the same number of internal states. They are therefore expected to find applications as quantum simulators of stochastic processes. Here, we emphasise that open quantum systems with instantaneous feedback are examples of HQMMs, thereby identifying a novel application of quantum feedback control. 
\keywords{Stochastic Processes, Hidden Markov Models, Quantum Simulations, Quantum Feedback.}
\end{abstract}

\section{Introduction}
In classical computer science, a Markov chain is a memoryless stochastic machine, which progresses from one state to another on a discrete time scale. Since their introduction in 1906 by Andrey Markov, the properties of Markov chains have been studied in great detail by mathematicians, computer scientists, and physicists alike \cite{Norris}. In the meantime, more complex versions of stochastic machines, like {\em Hidden Markov Models} (HMMs), have been introduced. These progress randomly from one internal state to another, which remains unobserved (hidden), while producing a stochastic output sequence. HMMs are widely used for the simulation of stochastic processes \cite{1,2,3}. Applications include speech recognition, image analysis and the modelling of biological systems.

Over recent years, several attempts have been made to extend the definition of HMMs into the quantum world and to utilise the properties of quantum systems to generate more complex stochastic output sequences \cite{4,5,6,HQMMAbstract,Petruccione}. For example, in 2011, Monras {\em et al.}~\cite{5} introduced so-called {\em Hidden Quantum Markov Models} (HQMMs). These are machines that progress from one quantum state to another, while generating classical output symbols. To produce an output symbol, a so-called generalised measurement or Kraus operation \cite{Kraus} is performed on the internal state of the machine.  One way of implementing a Kraus operation is to use an auxiliary quantum system --- a so-called ancilla. In every time step, the internal state of the HQMM interacts with its ancilla, which is then read out by a projective measurement. After every measurement, the ancilla is reset into its initial state, while the internal state of the HQMM remains hidden.  

A Kraus operation is the most general operation that a quantum system can experience, which is why Kraus operations are a vital part of the definition of a HQMM given by Monras {\em et al.}~\cite{5}. In a previous attempt to introduce quantum analogues of HMMs, Wiesner and Crutchfield \cite{4} defined so-called quantum finite-state generators, which only involved unitary operations and projective measurements. This is the most basic way in which a quantum system may be evolved and measured.  The state evolves according to the given unitary operator and is then measured, collapsing the state onto the measurement outcome.  In this way, they only obtained a subset of HQMMs, which are less powerful than their classical analogues. Different from quantum finite-state generators, HQMMs are able to produce more complex output sequences than HMMs with the same number of internal states.

Several ways of implementing HQMMs have already been identified:
\begin{enumerate}
\item As pointed out in Ref.~\cite{5}, one way of implementing HQMMs is the successive, non-adaptive read-out of entangled many-body states. 
\item Another example of a HQMM is the time evolution of an open quantum system on a coarse grained time scale, $\Delta t$, which produces a random sequence of classical output symbols. Indeed, in Ref.~\cite{Petruccione}, Sweke, Sinayskiy, and Petruccione use the language of HMMs to model open quantum systems.
\item The purpose of this paper is to highlight the connection between HQMMs and open quantum systems with {\em instantaneous feedback} \cite{Wiseman-Milburn}. In this way, we identify a way of implementing an even wider set of HQMMs. 
\end{enumerate}

Like HQMMs, open quantum systems evolve randomly in time. Taking this perspective, the open quantum system itself provides the internal states of a HQMM, while its surrounding bath plays the role of the ancilla, which is constantly reset into an environmentally preferred, or {\em einselected}, state \cite{Einselection}.  By this, we mean the state that the environment would naturally evolve into if left alone.  The continuous interaction between the internal states and the bath moves the bath away from its einselected state, thereby usually producing a measurable response that manifests itself as a random classical symbol. The effective dynamics of such a machine, when  averaged over all possible trajectories, can be described by a Markovian master equation \cite{Molmer,Hegerfeldt,Carmichael}. When describing an open quantum system in this way, its accompanying output sequence is ignored. Here we suggest not to do so and to use the output sequences of open quantum systems to simulate stochastic processes. Like HMMs, we expect HQMMs to find a wide range of applications \cite{5,7,Petruccione2}. 

Quantum feedback is a process in which the classical output symbols produced by an open quantum system are used to change its internal dynamics. Applications of quantum feedback control can be found, for example, in Quantum Information Processing \cite{Wiseman-Milburn}, where it is especially used to control state preparation \cite{StatePrep} and quantum transport \cite{Transport}. In these applications, the feedback is used to guide the internal dynamics of a quantum system. In contrast to this, this paper proposes to use quantum feedback to manipulate the classical output sequences of open quantum systems.  

There are five sections in this paper. Sect. \ref{HQMM} shortly reviews HQMMs. Afterwards, in Sect. \ref{open}, we review the master equations of Markovian open quantum systems with and without instantaneous feedback. Sect. \ref{continuous} shows that open quantum systems with instantaneous feedback are examples of HQMMs. Finally, we summarise our findings in Sect. \ref{conclusions}.

\section{Hidden Quantum Markov Models} \label{HQMM}

Hidden Markov Models (HMMs) are machines that evolve randomly from one internal state to another. In every time step, an output symbol is produced. Only the output symbol is detected externally, while the internal state of the machine remains hidden. Consequently, the time evolution of a HMM is characterised through a set of transition matrices $T_m$, where $m$ denotes the output symbol generated during the respective time step. For example, if the initial probability distribution of the internal states of the HMM is given by a vector $\vec{p}_0$, then the probability to obtain the outputs $abc \dots def$, where $a$ is the first symbol produced and $f$ is the last, is given by (see eg.~Ref.~\cite{5})
\begin{equation} \label{HMMexample}
p(abc\dots def) = \vec{\eta} \, T_f T_e T_d \dots T_c T_b T_a \, \vec{p}_0 \, .
\end{equation}
Here, $\vec{\eta}$ is a vector with all of its coefficients equal to 1.

Analogously, a Hidden Quantum Markov Model (HQMM) with a certain probability distribution of its internal state populations can be described by a density matrix, $\rho_{\rm S}$. In every time step, the system evolves and produces an output symbol.  Again, only the output symbol is detected externally, while the internal state of the machine remains hidden. In contrast to HMMs, the time evolution of a HQMM is governed by a set of Kraus operators $K_m$, where the subscripts $m$ coincide again with the output symbols of the machine. Using the same example as above, the probability of the output $abc \dots def$ occurring is now given by
\begin{equation} \label{HQMMexample}
p(abc\dots def) = {\rm Tr} \left(K_f K_e K_d \dots K_c K_b K_a \, \rho_{\rm S} \, K_a^{\dagger} K_b^{\dagger} K_c^{\dagger} \dots K_d^{\dagger} K_e^{\dagger} K_f^{\dagger} \right) \, .
\end{equation}
If the output symbol is ignored, then the density matrix $\rho_{\rm S}(t)$ of a HQMM evolves within a time step $(t,t+\Delta t)$ such that
\begin{equation} \label{HQMMKraus}
\rho_{\rm S}(t + \Delta t) = \sum\limits_{m=0}^{\infty} K_m \, \rho_{\rm S}(t) \, K_m^{\dagger} \, .
\end{equation}
The above Kraus operators $K_m$ should form a complete set. This means they need to obey the condition
\begin{equation} \label{Kraus}
\sum\limits_{m=0}^{\infty} K_m^{\dagger} K_m = 1 
\end{equation}
for the density matrix $\rho_{\rm S}(t + \Delta t)$ to be normalised. More details can be found in Ref.~\cite{5}.

\section{Open quantum systems} \label{open}

In the following, we describe how master equations can be used to model the time evolution of open quantum systems with linear couplings between the quantum system and its surrounding bath. Adopting the ideas of Zurek and others \cite{Einselection,Stokes}, we assume that the bath possesses an environmentally preferred state, a so-called einselected or pointer state. While the internal states of the open quantum system evolve on a relatively slow time scale, the bath relaxes rapidly back into its preferred state whenever its state is perturbed by the system-bath interaction. The more microscopic approach to the derivation of master equations, which we review here, makes it easy to incorporate instantaneous feedback into the dynamics of the open quantum system.

Our starting point is the Hamiltonian, $H$, for system and bath, which can be split into four parts, 
\begin{eqnarray} \label{H}
H &=& H_{\rm S} + H_{\rm int} + H_{\rm B} + H_{\rm SB} \, .
\end{eqnarray}
Here, $H_{\rm S}$ describes the free energy of the system and $H_{\rm int}$ allows for some internal system dynamics. Moreover, $H_{\rm B}$ represents the free energy of the bath and $H_{\rm SB}$ accounts for the system-bath interaction. When denoting the energy eigenstates of system and bath by $|n \rangle_{\rm S}$ and $|m \rangle_{\rm B}$, respectively, and assuming a linear coupling between system and bath, $H_{\rm S}$, $H_{\rm B}$, and $H_{\rm SB}$ can be written as
\begin{eqnarray} \label{System-Environment Hamiltonian}
H_{\rm S} & = & \sum\limits_{n=1}^N \hbar \omega_n \vert n \rangle_{\rm SS} \langle n \vert \, , \nonumber \\
H_{\rm B} & = & \sum\limits_{m=0}^{\infty} \, \hbar \omega_m \vert m \rangle_{\rm BB} \langle m \vert \, , \nonumber \\
H_{\rm SB} & = & \sum\limits_{m,m'=0}^{\infty} \sum\limits_{n,n' = 1}^N \hbar g_{nm, n'm'} \vert n'm' \rangle_{\rm SB \, SB} \langle nm \vert + \mbox{H.c.}  
\end{eqnarray}
without loss of generality. Because of being a bath, an infinite number of highly degenerate energy levels $\hbar \omega_m$ may occur. Finally, the $g$'s are system-bath coupling constants. Here, we assume for simplicity that these are time independent, though this is not always the case.

Since we are interested in identifying the relatively slow, effective internal dynamics of the open quantum system, we now move into the interaction picture with respect to the free system $H_0 =  H_{\rm B} + H_{\rm S}$, giving the interaction Hamiltonian
\begin{eqnarray} \label{Interaction Hamiltonian no feedback}
H_{\rm I} \left( t \right) &=& \sum\limits_{m,m'=0}^{\infty} \sum\limits_{n,n' = 1}^N \hbar g_{nm,n'm'} \vert n'm' \rangle_{\rm SB \, SB} \langle nm \vert \, {\rm e}^{ -{\rm i} \left(\omega_m - \omega_{m'} + \omega_n - \omega_{n'} \right) t} \nonumber \\
&& + \mbox{H.c.} + H_{\rm int\, I} \left(t\right) 
\end{eqnarray}
with $H_{\rm int\, I} \left(t\right) $ describing the internal dynamics of the system in the interaction picture. The Hamiltonian $H_{\rm I}$ no longer contains free energy terms. The time evolution of system and environment in the interaction picture is hence much slower than in the Schr\"odinger picture.

Suppose the environment, which the bath couples to, thermalises very rapidly, thereby relaxing the bath into an environmentally preferred state --- a so-called pointer state. This state minimises the entropy of the bath and does not evolve in time unless there is a very strong system-bath interaction. In the following, we denote the corresponding bath state by $|0 \rangle_{\rm B}$. Without restrictions, the above introduced notation can indeed be chosen such that the pointer state corresponds to $m=0$. Moreover, we can choose the free energy of the pointer state such that $\omega_{0} = 0$, again without loss of generality.

Next we assume that the initial state of the open quantum system at a time $t$ is given by the density matrix $\rho_{\rm S} \left(t \right)$, while the bath is in $|0 \rangle_{\rm B}$. Over a short time $\Delta t$, the system and bath then evolve in the interaction picture into the density matrix $\rho_{\rm SB} (t+\Delta t)$ with
\begin{equation} \label{Evolution2}
\rho_{\rm SB} (t+\Delta t) = U_I\left(t+\Delta t , t\right) \vert 0 \rangle_{\rm B} \, \rho_{\rm S}(t) \, _{\rm B} \langle 0 \vert U_I^{\dagger} \left(t+\Delta t , t\right) \, .
\end{equation}
Subsequently, on a time scale that is fast compared to $\Delta t$, the surrounding bath thermalises again, which transforms it back into its pointer state. Due to locality, this process should only affect the bath and not the quantum system itself. All the expectation values of the system should remain the same during the relaxation process. Consequently, the time evolution of system and bath within $\Delta t$ can be summarised as
\begin{equation} \label{Evolution3}
\rho_{\rm SB} (t+\Delta t) \longrightarrow \vert 0 \rangle_{\rm B}  \rho_{\rm S}(t+\Delta t) _{\rm B} \langle 0 \vert 
\end{equation}
with
\begin{equation} \label{General evolution no feedback}
\rho_{\rm S}\left(t+\Delta t\right) = {\rm Tr}_{\rm B} \left( \rho_{\rm SB} (t+\Delta t) \right) \, .
\end{equation}
This equation describes an effective Markovian system dynamics within the time interval $(t,t+\Delta t)$. To summarise the effective time evolution of the open quantum system in a more compact form, i.e.~in form of a master equation, we now calculate the time derivative of $\rho_{\rm S}(t)$, 
\begin{equation} \label{Derivative}
\dot{\rho}_{\rm S} = \lim_{\Delta t \to 0} {1 \over \Delta t} \left( \rho_{\rm S} \left(t+ \Delta t \right) - \rho_{\rm S}\left( t \right) \right) \, .
\end{equation}
Using the above mentioned time scale separation between the internal and external dynamics of the open quantum system allows us to evaluate equation~(\ref{General evolution no feedback}) using second order perturbation theory, which implies
\begin{eqnarray} \label{perturbation}
U_{\rm I} (\Delta t ,0) &=& 1 - \frac{i}{\hbar} \int_0^{\Delta t} {\rm d}t \, H_{\rm I}\left(t \right) 
- \frac{1}{\hbar^2} \int_0^{\Delta t} {\rm d}t \int_0^{t} {\rm d}t' \, H_{\rm I}\left(t \right) H_{\rm I} \left(t' \right) \, .
\end{eqnarray}
Substituting Eqs.~(\ref{Interaction Hamiltonian no feedback}), (\ref{General evolution no feedback}) and (\ref{perturbation}) into equation~(\ref{Derivative}) eventually yields a master equation of the general form (see e.g.~Refs.~\cite{Molmer,Hegerfeldt,Carmichael,Stokes} for more details)
\begin{eqnarray} \label{Lindblad}
\dot{\rho}_{\rm S} &=& - \frac{i}{\hbar} \left[ H_{\rm int \, I} , \rho_{\rm S} \right] 
- \frac{1}{2} \sum\limits_{n,n',n'',n''' =1}^{N} \xi_{nn'} \xi_{n''n'''}^* \, \left[L_{n''n'''}^{\dagger} L_{nn'} , \rho_{\rm S} \right]_+ \nonumber \\
&& + \sum\limits_{n,n',n'',n''' =1}^{N} \xi_{nn'} \xi_{n''n'''}^* \, L_{nn'} \, \rho_{\rm S} \, L_{n''n'''}^{\dagger} \, . ~~
\end{eqnarray} 
The $L$'s in this equation are operators that act on the internal states of the open quantum system and the $\xi$'s are constants. In addition, one can show that the above equation is of Lindblad form \cite{Lindblad}, which is the most general way of expressing the master equation for a Markovian quantum system.  

Equation~(\ref{Lindblad}) describes open quantum systems without feedback. These are quantum systems, where the environment does nothing else but constantly resets the bath that surrounds the system back into its pointer state. However, this is not necessarily the case. Open quantum systems can be designed such that the population of a state $|m \rangle_{\rm B}$ that is {\em not} environmentally preferred, triggers a back action, which changes the density matrix $\rho_{\rm S}(t)$ by a certain unitary operation $R_m$. Such a back action is known as {\em feedback}. If the feedback is so fast such that its time scale is short compared to the time scale on which $\rho_{\rm S}(t)$ evolves, then we talk about instantaneous feedback. Using the same arguments as above, one can show that the open quantum system evolves in this case according to a master equation of the general form
\begin{eqnarray} \label{Lindblad2}
\dot{\rho}_{\rm S} &=& - \frac{i}{\hbar} \left[ H_{\rm int \, I} , \rho_{\rm S} \right] \nonumber \\
&& - \frac{1}{2}  \sum_{m=1}^\infty \sum\limits_{n,n',n'',n''' =1}^{N} \xi_{nn',m} \xi_{n''n''',m}^* \, \left[L_{n''n''',m}^{\dagger} L_{nn',m} , \rho_{\rm S} \right]_+ \nonumber \\
&& + \sum_{m=1}^\infty \sum\limits_{n,n',n'',n''' =1}^{N} \xi_{nn',m} \xi_{n''n''',m}^* \, L_{nn',m} \, \rho_{\rm S} \, L_{n''n''',m}^{\dagger}  
\end{eqnarray}
with the $L_{nn',m}$ operators defined such that
\begin{equation}
L_{nn',m} = R_m \, L_{nn'} \, .
\end{equation}
This equation is of exactly the same form as the master equation for open quantum systems with instantaneous feedback in Ref.~\cite{Wiseman-Milburn}. 

\section{Open quantum systems as HQMMs} \label{continuous}

Comparing the above description of open quantum systems with the definition of the HQMM in Sect. \ref{HQMM}, it becomes relatively straightforward to see that open quantum systems with instantaneous feedback are concrete examples of HQMMs. To illustrate this in more detail, we notice that $\rho_{\rm S} (t+\Delta t)$ in equation~(\ref{General evolution no feedback}) is a statistical mixture of different subensembles and distinguish two cases. 

\subsection{Energy exchange with bath and environment} 

The first case is the one, where the bath has been reset into its pointer state, $|0 \rangle_{\rm B}$, within $(t,t+\Delta t)$ after having evolved into $|m \rangle_{\rm B}$ and experiencing the feedback operation $R_m$. The above equations and their given interpretation tell us that the density matrix of the corresponding subensemble equals
\begin{equation} \label{Kraus emission}
\rho_{\rm S}(t+\Delta t |m \ge 1) = K_m \, \rho_{\rm S}(t) \, K_m^\dagger
\end{equation}
in this case, with the operator $K_m$ given by
\begin{equation} \label{Km}
K_m = \sum\limits_{n,n' =1}^{N} \xi_{nn',m} \, L_{nn'm} \sqrt{\Delta t}
\end{equation}
for $m \ge 1$. As we shall see below, $K_m$ is a Kraus operator, which acts on the internal state of the open quantum system.

\subsection{No energy exchange between bath and environment}

The remaining terms in the above master equation correspond to $m=0$ and describe the time evolution of the open quantum system under the condition that the surrounding bath remains in its environmentally preferred state $|0 \rangle_{\rm B}$. In this case, $\rho_{\rm S}(t)$ evolves within $\Delta t$ into
\begin{equation} \label{Kraus no energy}
\rho_{\rm S} (t+\Delta t |m=0) = K_0 \, \rho_{\rm S}(t) \, K_0^\dagger \, .
\end{equation}
Up to first order in $\Delta t$, the corresponding operator $K_0$ can be written as
\begin{equation} \label{K0}
K_0 = {\rm exp} \left( - \frac{{\rm i}}{\hbar} H_{\rm cond} \Delta t \right) 
\end{equation}
with the non-Hermitian Hamiltonian $H_{\rm cond}$ given by
\begin{equation}
H_{\rm cond} = H_{\rm int \,I} - \frac{{\rm i}}{2} \hbar \sum_{m=1}^\infty \sum\limits_{n,n',n'',n''' =1}^{N} \xi_{nn',m} \xi_{n''n''',m}^* \,  L_{n''n''',m}^{\dagger} L_{nn',m} \, .
\end{equation}
The last term in this equation is crucial for the density matrix $\rho_{\rm S} (t+\Delta t)$ in equation~(\ref{Lindblad2}) to remain normalised.

\subsection{Comparison of Kraus operators}

To show that open quantum systems with instantaneous feedback are examples of HQMMs, we now only need to identify the operators $K_m$ in Eqs.~(\ref{Km}) and (\ref{K0}) with the Kraus operators in equation~(\ref{HQMMKraus}). Summing over all of the above described subensembles with their respective output symbols given by $m=0,1,...$, we immediately see that equation~(\ref{HQMMKraus}) applies. Since a density matrix $\rho_{\rm S}(t)$, which evolves according to the master equation of an open quantum system in Lindblad form remains normalised, we moreover have
\begin{equation} \label{Kraus check}
{\rm Tr}_{\rm S} \left( \sum\limits_{m=0}^{\infty} K_m \, \rho_{\rm S} \, K_m^{\dagger} \right) 
= {\rm Tr}_{\rm S} \left( \sum\limits_{m=0}^{\infty} K_m^{\dagger} K_m \, \rho_{\rm S} \right) = 1 \, .
\end{equation}
This means, equation~(\ref{Kraus}) too is satisfied. Open quantum systems with instantaneous feedback are indeed examples of HQMMs.

\section{Conclusions} \label{conclusions}

Motivated by the popularity of Hidden Markov Models (HMMs) in classical computer science, this paper has a closer look at the quantum analogues of these machines --- so-called Hidden Quantum Markov Models (HQMMs) \cite{5}. Sect.~\ref{HQMM} defines HQMMs in terms of Kraus operators. Sect.~\ref{open} gives an overview of how to model open quantum systems with and without instantaneous feedback with the help of master equations in Lindblad form \cite{Wiseman-Milburn,Molmer,Hegerfeldt,Carmichael,Lindblad}. When comparing Sects.~\ref{HQMM} and \ref{open} in Sect.~\ref{continuous}, it becomes obvious open quantum systems with random classical output sequences are examples of HQMMs. This paper proposes not to ignore the random classical output sequences of open quantum systems, since they could find interesting applications as quantum simulators of stochastic processes.  

Finally, it might be worth noting that the above analysis of open quantum systems with instantaneous feedback only allows for an environmental back action when the system-bath interaction changes the bath into a state that is different from its environmentally preferred state, $|0 \rangle_{\rm B}$. This need not be the case. Physically, it is possible to design open quantum systems, which experience feedback also, when no exchange of energy occurs between system and bath. In this case, the open quantum system can no longer be modelled by a master equation. However, the effective system dynamics would remain Markovian and could be described using the language of HQMMs. \\[0.5cm]
{\em Acknowledgement.} T.~B.~acknowledges financial support from a White Rose Studentship Network on Optimising Quantum Processes and Quantum Devices for future Digital Economy.


\begin{thebibliography}{Dirac 58}
\bibitem{Norris}
J. R. Norris, {\em Markov chains}, Cambridge University Press (1998).

\bibitem{1} 
L. R. Rabiner, {\em A tutorial on hidden Markov models and selected applications in speech recognition}, Proc. IEEE {\bf 77}, 257 (1989).

\bibitem{2} 
H. Xue, {\em Hidden Markov Models Combining Discrete Symbols and Continuous Attributes in Handwriting Recognition}, IEEE Transactions on Pattern Analysis and Machine Intelligence {\bf 28}, 458 (2006).

\bibitem{3}
B. Vanluyten, J. C. Willems, and B. D. Moor, {\em Equivalence of State Representations for Hidden Markov Models}, Systems and Control Letters {\bf 57}, 410 (2008).

\bibitem{4}
K. Wiesner and C. P. Crutchfield, {\em Computation in finitary stochastic and quantum processes}, Physica D {\bf 237}, 1173 (2008).

\bibitem{5} 
A. Monras, A. Beige, and K. Wiesner, {\em  Hidden Quantum Markov Models and non-adaptive read-out of many-body states}, App. Math. and Comp. Sciences {\bf 3}, 93 (2011).

\bibitem{6} 
P. Gmeiner, {\em Equality conditions for internal entropies of certain classical and quantum models},
arXiv:1108.5303 (2011).

\bibitem{HQMMAbstract} 
B. O`Neill, T. M. Barlow, D. Safranek, and A. Beige, {\em  Hidden Quantum Markov Models with one qubit}, AIP Conf. Proc. {\bf 1479}, 667 (2012).

\bibitem{Petruccione}
R. Sweke, I. Sinayskiy, and F. Petruccione, {\em Simulation of Single-Qubit Open Quantum Systems}, arXiv:1405.6049 (2014).

\bibitem{Kraus}
K. Kraus, {\em States, Effects and Operations},  Lecture Notes in Physics {\bf 190} (Springer-Verlag, Berlin, 1983).

\bibitem{Wiseman-Milburn} 
H. M. Wiseman and G. J. Milburn, {\em Quantum Measurement and Control}, Cambridge University Press (2010).

\bibitem{Einselection} 
W. H. Zurek, {\em  Decoherence, einselection, and the quantum origins of the classical}, Rev. Mod. Phys. {\bf 75}, 715 (2003).

\bibitem{Molmer}
J. Dalibard, Y. Castin, and K. M{\/o}lmer, {\em Wave-function approach to dissipative processes in quantum optics} Phys. Rev. Lett. {\bf 68}, 580 (1992).

\bibitem{Hegerfeldt}
G. C. Hegerfeldt, {\em How to reset an atom after a photon detection. Applications to photon counting processes}, Phys. Rev. A {\bf 47}, 449 (1993).

\bibitem{Carmichael}
H. Carmichael, {\em An Open Systems Approach to Quantum Optics},  Lecture Notes in Physics {m18} (Springer-Verlag, Berlin, 1993).

\bibitem{7}
N. Goldenfeld and C. Woese, {\em Life is Physics: evolution as a collective phenomenon far from equilibrium}, Ann. Rev. Cond. Matt. Phys. {\bf 2}, 375 (2011).

\bibitem{Petruccione2}
M. Schuld, I. Sinayskiy, and F. Petruccione, {\em Quantum walks on graphs representing the firing patterns of a quantum neural network},  arXiv:1404.0159 (2014).

\bibitem{StatePrep} 
G. Kiesslich, C. Emary, G. Schaller, and T. Brandes, {\em Reverse quantum state engineering using electronic feedback loops}, New. J. Phys. {\bf 14}, 123036 (2012).

\bibitem{Transport} 
C. Emary, {\em Delayed feedback control in quantum transport}, Phil. Trans. R. Soc. A {\bf 371}, 1999 (2013).

\bibitem{Stokes}
A. Stokes, A. Kurcz, T. P. Spiller, and A. Beige, {\em Extending the validity range of quantum optical master equations}, Phys. Rev. A {\bf 85}, 053805 (2012).

\bibitem{Lindblad} 
G. Lindblad, {\em On the generators of quantum dynamical semigroups}, Comm. Math. Phys. {\bf 48}, 119 (1976).
\end{thebibliography}
\end{document}